\documentclass[12pt]{article}

\usepackage[utf8]{inputenc}      
\usepackage[T1]{fontenc}         
\usepackage{lmodern}             

\usepackage{amsmath,amssymb,amsthm}

\usepackage{graphicx}
\usepackage{siunitx}             
\DeclareSIUnit\bar{bar}
\usepackage[a4paper,margin=1in]{geometry}

\usepackage{float}               
\usepackage{caption}
\usepackage{booktabs}            

\usepackage[%
  backend=biber,
  style=ieee,
  citestyle=numeric-comp,   
  sorting=none,
  doi=false,isbn=false,url=false
]{biblatex}
\usepackage{hyperref}
\hypersetup{
  colorlinks=true,
  linkcolor=blue,
  citecolor=blue,
  urlcolor=blue,
  pdfpagemode=UseNone
}
\usepackage[capitalize,nameinlink]{cleveref}
\crefname{figure}{Fig.}{Figs.}
\crefformat{figure}{#2Fig.~#1#3}

\usepackage{authblk}

\usepackage{todonotes}

\newcommand{\GaAs}{\ensuremath{\mathrm{GaAs} \,}}

\newcommand{\AlGaAs}{\ensuremath{\mathrm{Al}_{0.15}\mathrm{Ga}_{0.85}\mathrm{As} \,}}
\newcommand{\nm}{\nano\metre}

\newcommand{\mm}{\milli\metre}
\newcommand{\EEfiber}{\ensuremath{\text{EE}^\text{fiber}\,}}
\newcommand{\EEfiberNom}{\ensuremath{\text{EE}^\text{fiber}_{NA \to 0.6}\,}}
\newcommand{\EEfree}{\ensuremath{\text{EE}^\text{free}\,}}
\newcommand{\EEfreeMax}{\ensuremath{\text{EE}^\text{free}_{NA \to 1}\,}}
\newcommand{\EEfreeNom}{\ensuremath{\text{EE}^\text{free}_{NA \to 0.6}\,}}

\DeclareUnicodeCharacter{0308}{{\"o}}   
\DeclareUnicodeCharacter{2009}{ }       

\title{Free-standing circular Bragg gratings enabling efficient GaAs quantum dot entangled photon pair sources}

\author[1]{Sai Abhishikth Dhurjati}
\author[1]{Moritz Langer}
\author[1]{Yared G. Zena}
\author[1]{Ahmad Rahimi}
\author[1]{Liesa Raith}
\author[1]{Martin Bauer}
\author[3]{Frank H. P. Fitzek}
\author[2]{Riccardo Bassoli}
\author[2]{Caspar Hopfmann\thanks{Corresponding author, email: caspar\_arndt.hopfmann@tu-dresden.de}}

\affil[1]{Institute for Emerging Electronic Technologies, IFW Dresden, Helmholtzstraße 20, 01069 Dresden, Germany}
\affil[2]{Quantum Communication Networks research group, Deutsche Telekom Chair of Communication Networks, Dresden University of Technology, Germany}
\affil[3]{Deutsche Telekom Chair of Communication Networks, Dresden University of Technology, Germany}

\begin{document}

\date{\today}

\maketitle

\begin{abstract}

Deterministic and bright quantum light sources based on scalable semiconductor technologies are a crucial building block for future quantum communication networks. While circular Bragg gratings (CBGs) are highly effective for extracting light from solid-state quantum emitters, conventional architectures rely on complex multi-layer processing or flip-chip bonding, which introduce detrimental strain and limit scalability. Here, we present a fabrication-minimal approach to realize monolithic, free-standing CBG cavities with deterministically positioned single GaAs quantum dots (QDs). By utilizing aspect-ratio-dependent etching (ARDE) in a single-step top-down process, we achieve the necessary vertical structural asymmetry for directional emission without requiring bottom reflectors. Finite-difference time-domain (FDTD) simulations validate this geometry, predicting free-space extraction efficiencies up to \qty{68}{\percent} and coupling efficiencies of \qty{40}{\percent} into a lensed single-mode fiber (\text{NA} = 0.6). Experimentally, the deterministically coupled QD-CBG devices yield a photoluminescence intensity enhancement of up to $\times \num{700}$ comp ared to unprocessed planar QDs, reaching integrated count rates of \qty{45}{\mega\hertz}. Furthermore, the suspended membrane architecture effectively relaxes residual strain, significantly reducing the average exciton fine-structure splitting from \qty{7.3}{\micro\eV} in planar QDs to \qty{1.3}{\micro\eV} in the CBGs. Interferometric measurements confirm that the fabrication process preserves the optical quality of the emitters, with average coherence times of \qty{70}{\pico\second}. By bridging optimized FDTD design with precise nanofabrication and robust optical performance, these results establish free-standing GaAs CBGs as a highly scalable platform for bright and coherent entangled photon pair sources.
\end{abstract}

\section{Introduction}

In the quest to realize scalable quantum communication and quantum-network infrastructures, significant progress has been achieved through the development of high-performance solid-state quantum light sources \cite{Liu2019, Claudon2010, Dousse2010, Hopfmann2021}. As these technologies transition from laboratory demonstrations toward practical deployment, increasing emphasis is placed on compactness, fabrication simplicity, long-term stability, and compatibility with fiber-based architectures operating at cryogenic temperatures \cite{Schlehahn2018, Musial2020, Langer2025, Rickert2025}. In this context, photonic platforms that enable efficient and reproducible photon extraction from semiconductor quantum emitters while maintaining a minimal system footprint are of central importance \cite{Langer2025, Bremer2020, Zena2026}. Among the various solid-state quantum emitters explored to date, semiconductor quantum dots (QDs) grown by molecular beam epitaxy (MBE) offer discrete energy levels, excellent optical quality, and full compatibility with wafer-scale fabrication. In particular, local-droplet-etched GaAs/AlGaAs QDs have emerged as a premier platform for polarization-entangled photon pair sources due to their inherently high spatial symmetry and near-zero excitonic fine-structure splitting (FSS) \cite{Huber2017}. However, the practical usability of such emitters in native planar heterostructures is fundamentally limited by inefficient photon extraction caused by bulk leakage, total internal reflection at high-index interfaces, and non-directional emission \cite{Liu2019, Langer2025a}.
To address these challenges, a variety of photonic microstructures have been developed, including nanopillars, solid-immersion lenses, and microlenses \cite{Tomm2021, Dousse2010, Hopfmann2021, Langer2025a, Liu2019}. While high-Q microcavities offer strong Purcell enhancement, their narrow spectral bandwidths complicate the simultaneous enhancement of both the exciton and biexciton transitions -- a strict requirement for efficient entangled photon pair generation \cite{Dousse2010}. Consequently, circular Bragg gratings (CBGs) have emerged as a highly attractive alternative. CBGs provide moderate optical confinement, strong Purcell enhancement, and highly directional vertical emission over a broad spectral bandwidth, overcoming the stringent spectral alignment constraints of traditional microcavities \cite{Liu2019}.
Despite their optical advantages, most reported CBG implementations rely on additional bottom reflectors, such as metallic mirrors or dielectric stacks, to ensure unidirectional upward emission \cite{Liu2019, Wang2019, MoczalaDusanowska2020, Holewa2024}. Fabricating these mirror-backed architectures typically requires flip-chip bonding or complex multilayer processing. These highly invasive top-down workflows are known to introduce microscopic strain, crystal defects, and process-induced inhomogeneities \cite{Zhai2020}. For entangled photon pair sources, such mechanical strain is highly detrimental: it breaks the epitaxial symmetry of the QD, drastically increasing the FSS and leading to exciton precession oscillations which degrades the observability of entanglement \cite{Winik2017, Hopfmann2021}. Furthermore, multi-step alignment processes limit scalability and reduce overall device yield. Consequently, there is strong motivation to develop simpler, strain-free cavity platforms that preserve both high optical performance and native emitter properties while remaining compatible with scalable fabrication.
In this work, we investigate a platform based on vertically asymmetric, free-standing GaAs CBGs integrated with deterministically positioned GaAs QDs. Inspired by suspended grating concepts \cite{Davanco2011, Rickert2023}, we employ an intentional vertical structural asymmetry—achieved via partial trench etching—in a suspended membrane geometry. This effectively negates the need for bottom reflectors while maintaining broadband cavity operation and highly efficient vertical extraction. Crucially, the entire structure is defined in a single-step lithography and pattern-transfer process that exploits aspect-ratio-dependent etching (ARDE) to simultaneously create shallow grating trenches and deep under-etch release holes. To ensure high device yield and reproducible emitter–cavity coupling across the entire ensemble, all devices are spatially aligned to pre-selected QDs using atomic force microscope (AFM) nano-oxidation lithography, as detailed in our recent work \cite{Dhurjati2026}. We perform comprehensive finite-difference time-domain (FDTD) simulations to bridge the gap between ideal design and realized geometries, followed by a rigorous optical and structural characterization of the fabricated devices. By demonstrating massive intensity enhancements, preserved coherence, and effectively relaxed strain (yielding a heavily reduced FSS compared to bulk), our results establish free-standing GaAs CBGs as a robust, scalable, and fabrication-tolerant platform for integrated quantum photonics.

\section{Concept and Device Design}
\label{sec:concept}

\begin{figure}[H]
    \centering
    \includegraphics[width=\columnwidth]{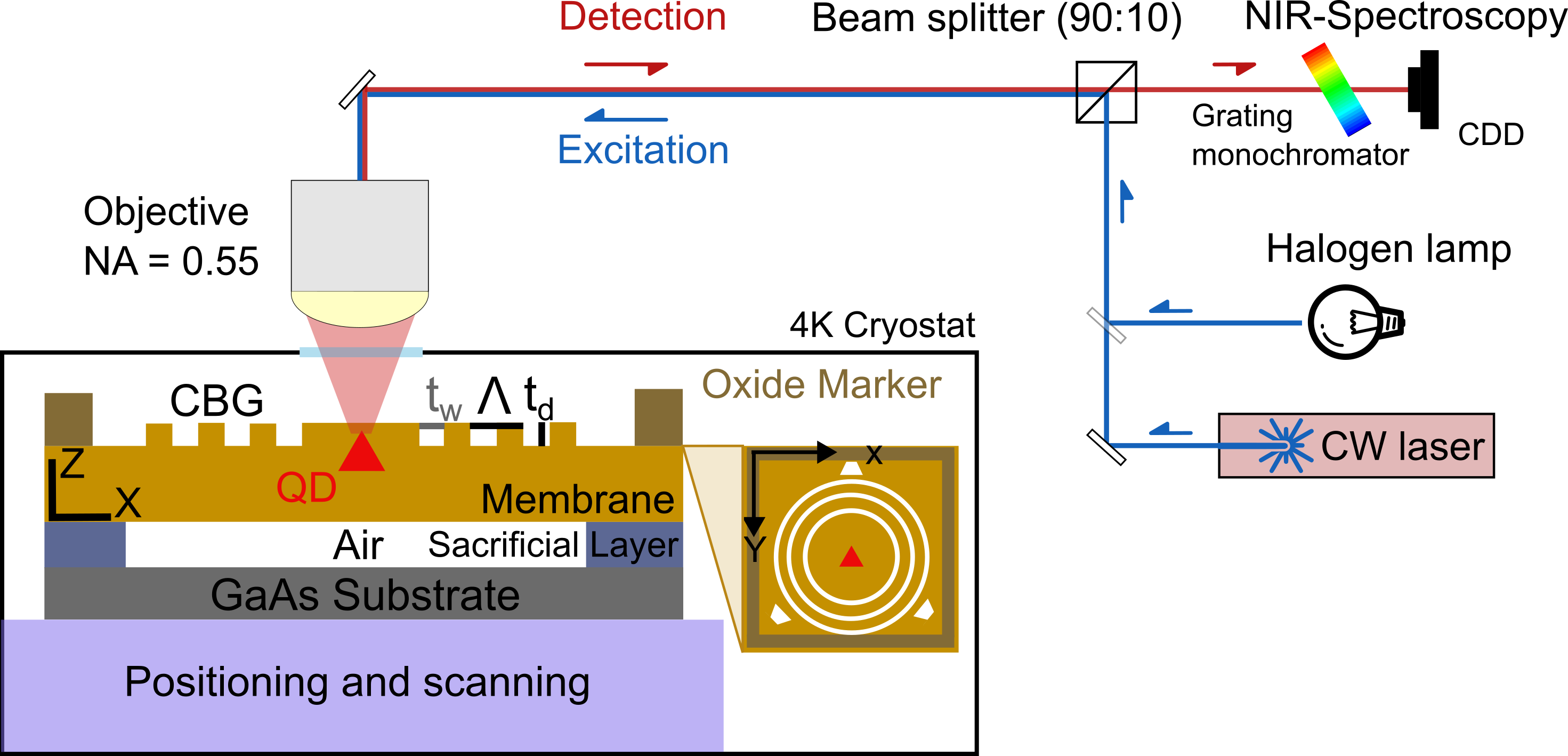}
    \caption{\label{fig:Setup_Fabrication}
    Schematic overview of the ultra-compact fiber-coupled photon source placed within a \qty{4}{K} cryostat and the experimental characterization apparatus. The free-space collection objective features a numerical aperture of \num{0.55}. A \qty{635}{\nm} continuous wave laser is used for excitation within the near-infrared (NIR) $\mu$-photoluminescence ($\mu$PL) spectroscopy, while a halogen lamp is employed for broadband reflection spectroscopy.
    }
\end{figure}
                
To realize compact and efficient solid-state quantum light sources compatible with scalable photonic integration, we employ QDs grown by local droplet etching (LDE) \cite{Sanguinetti2003, Heyn2007, Keil2017} using MBE. Spontaneous emission from QDs embedded in bulk high-index semiconductor materials, such as \AlGaAs{}, is intrinsically inefficient due to total internal reflection at the semiconductor-air interfaces. Combined with the isotropic nature of dipole radiation, the fraction of photons recoverable by external collection optics is typically limited to a few percent \cite{Lodahl2015}. To overcome these extraction limitations, we integrate the QDs into circular Bragg grating (CBG) cavities. CBGs uniquely combine moderate optical confinement and broadband spectral operation with highly directional vertical emission \cite{Ates2012, Liu2019}. 

In contrast to high-Q microcavities, which offer strong Purcell enhancement ($F_p$) but suffer from narrow spectral bandwidths and stringent emitter-cavity spectral resonance requirements \cite{Andreani1999}, CBGs operate in a moderate-Q regime. This broader spectral bandwidth is a critical requirement for entangled photon pair sources, as it allows for the simultaneous enhancement of both the exciton (X) and biexciton (XX) transitions. Furthermore, the moderate reduction in radiative lifetime shortens the emission window, suppresses background emission, and reduces susceptibility to environmental dephasing \cite{Somaschi2016}, all without the extreme fabrication tolerances demanded by high-Q microcavities.

A key design innovation of the present platform is the use of vertically asymmetric, free-standing CBGs fabricated within suspended \AlGaAs{} membranes. Previously reported CBG architectures typically achieve unidirectional upward emission by incorporating bottom reflectors, such as metallic mirrors \cite{Liu2019, Holewa2024} or hybrid dielectric-metal stacks \cite{Barbiero2022}. These approaches often require flip-chip bonding or complex multilayer deposition. While optically effective, these invasive processing steps can induce mechanical strain, crystal defects, or epitaxial symmetry breaking in the QD layer. Such effects degrade optical coherence and increase the excitonic fine-structure splitting (FSS)---a severely detrimental outcome for polarization-entangled photon pair generation \cite{Michler2014, Hopfmann2021, Yang2022}. In our free-standing membrane geometry, preferential upward emission is instead achieved by the incomplete vertical etching of the circular trenches, as illustrated in \cref{fig:Setup_Fabrication}. This partial etch creates a vertical asymmetry in the CBG membrane cross-section, directing light upward analogously to a bottom reflector, but entirely avoiding strain-inducing bonding or deposition steps \cite{Davanco2011}. 

To optimize this architecture, all CBG geometries are designed using three-dimensional finite-difference time-domain (FDTD) simulations targeting the typical \GaAs{} QD emission wavelength of \qty{780}{\nm}. The simulations incorporate the temperature- and wavelength-dependent refractive indices of \AlGaAs{} previously determined by our group \cite{Langer2025b}. As defined in \cref{fig:Setup_Fabrication}, the key structural parameters are the suspended membrane thickness $t_M$, trench width $t_W$, grating period $\Lambda$ (center-to-center spacing between adjacent trenches), trench depth $t_d$, and the number of grating rings $N_r$. The model assumes a centrally located in-plane dipole emitter embedded within a central mesa of radius $2\Lambda$. This enlarged central mesa is deliberately chosen to maximize the distance between the QD and the first etched trench, thereby protecting the emitter from non-radiative surface defects induced by reactive-ion etching. The parameter $N_r$ is chosen to balance cavity $Q$-factor against spectral bandwidth \cite{Ates2012} in order to accommodate the X-XX cascade and fabrication-induced spectral detuning, the target $Q$-factor is constrained to $\leq \num{300}$. The primary optimization figure of merit is the extraction efficiency of the QD into the guided mode of a lensed single-mode fiber with a numerical aperture (NA) of \num{0.6}, denoted as \EEfiberNom. This metric is calculated by overlapping the free-space far-field emission (\EEfree) with the mode profile of the lensed fiber \cite{Nie2021, Langer2025}.

\begin{figure}[H]
    \centering
    \includegraphics[width=0.8\textwidth]{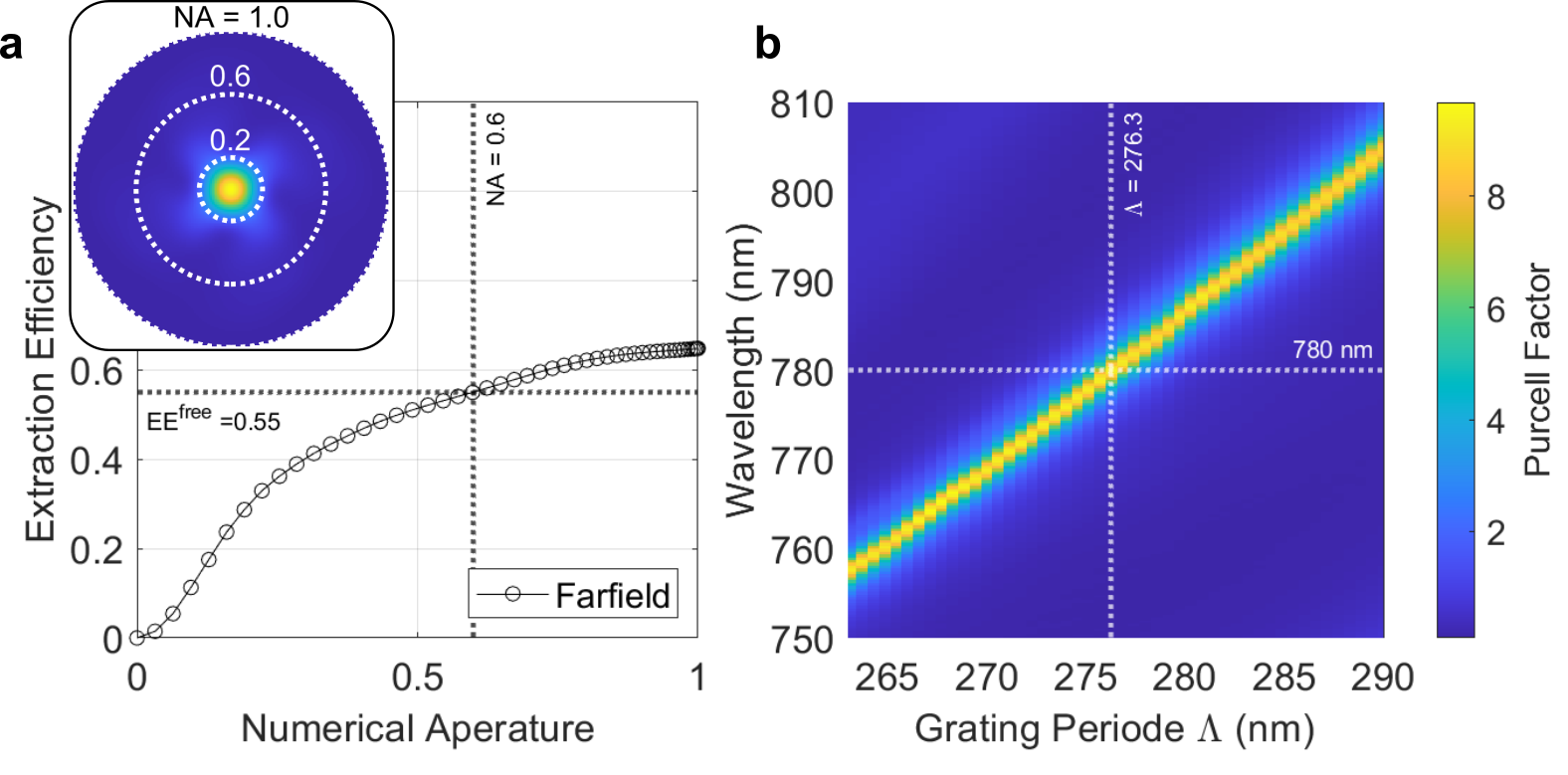}
    \caption{\label{fig:Ideal_FDTD Simulations}
    Simulated device performance of an optimized monolithic, free-standing CBG. The device parameters are listed in \cref{tab:CBG_parameters}. The optimization figure of merit is \EEfiberNom. (a) Free-space extraction efficiency as a function of the collection numerical aperture (NA); the maximal obtained value at an NA of \num{0.6} is \num{0.55}. Inset: Squared electrical field amplitude of the CBG far-field emission in polar representation. (b) Purcell factor of the dipole source versus its emission wavelength and grating period $\Lambda$.}
\end{figure}

The resulting FDTD-optimized parameters for the ideal geometry are a membrane thickness $t_M = \qty{134}{\nm}$, trench width $t_W = \qty{77}{\nm}$, grating period $\Lambda = \qty{276}{\nm}$, trench depth $t_d = \qty{83}{\nm}$, and $N_r = \num{6}$ rings (\cref{tab:CBG_parameters}). In this ideal configuration, the trenches are etched to a depth of roughly \qty{62}{\percent} of the membrane thickness. This shallow-etch profile preserves mechanical integrity, minimizes sidewall scattering, and provides the requisite refractive-index contrast for the grating resonance. Simulations for this ideal design predict a \EEfiberNom and a maximum free-space extraction \EEfreeMax of \qty{40}{\percent} and \qty{68}{\percent}, respectively. As shown in \cref{fig:Ideal_FDTD Simulations}a, the free-space efficiency evaluated at an NA of \num{0.6} (\EEfreeNom) reaches \qty{55}{\percent}. This design yields a cavity Q-factor of \num{212} and a Purcell factor of $F_p = \num{9.4}$. The spectral tuning of the fundamental CBG resonance and $F_p$ as a function of the grating period $\Lambda$ is illustrated in \cref{fig:Ideal_FDTD Simulations}b.

Due to technical constraints in molecular beam epitaxy, starting material with the exact ideal membrane thickness of \qty{134}{\nm} was unavailable. Consequently, \qty{150}{\nm}-thick \AlGaAs{} membranes are utilized for device fabrication. The CBG parameters were rigorously re-optimized for this specific thickness, yielding updated target values of $t_W = \qty{83}{\nm}$, $\Lambda = \qty{270}{\nm}$, and $t_d = \qty{103}{\nm}$, which corresponds to a partial etch of approximately \qty{68}{\percent} of the membrane thickness (\cref{tab:CBG_parameters}). For this practically realizable geometry, simulations predict \EEfiberNom and \EEfreeNom of \qty{31}{\percent} and \qty{50}{\percent}, respectively, alongside a Q-factor of \num{265} and $F_p = \num{11.1}$. The wavelength dependence and NA-dependent efficiencies for this structure are detailed in the supplementary material. Deviations from these targeted geometries arising from fabrication tolerances are quantified and discussed in \cref{sec:fabrication}.

Finally, while robust cavity design is paramount, random spatial misalignment between the QD and the cavity center drastically degrades extraction efficiency and polarization properties. To ensure high-yield, reproducible emitter–cavity coupling across the device ensemble, all fabricated CBGs are deterministically aligned to pre-selected QDs using atomic force microscope nano-oxidation lithography (AFM-NL) \cite{Cambel2007, Cambel2008}. By utilizing localized AFM-written oxide markers as high-contrast alignment references for electron beam lithography, the cavity center is precisely placed on a buried QD. A detailed characterization of this positioning technique is provided in our recent work \cite{Dhurjati2026}. Here, AFM-NL serves as the crucial enabling step that translates optimized FDTD designs into functional, high-performance experimental devices.

\section{Device Fabrication}
\label{sec:fabrication}

Free-standing CBG cavities deterministically aligned to selected QDs are realized using a highly streamlined, single top-down nanofabrication sequence. Crucially, all pattern elements--including both the concentric grating trenches and the outer under-etch release holes--are defined in a single electron beam lithography (EBL) step. This minimizes the total number of processing steps and entirely eliminates the overlay errors inherent to multi-step lithography. The starting material is a planar heterostructure grown by MBE, consisting of a \qty{150}{\nm}-thick \AlGaAs{} membrane situated above a \qty{250}{\nm}-thick $\mathrm{Al}_{0.75}\mathrm{Ga}_{0.25}\mathrm{As}$ sacrificial layer on a \GaAs{} substrate. The QD layer is embedded at the vertical center of the \AlGaAs{} membrane (a depth of \qty{75}{\nm}), defined by the LDE growth conditions. Scanning electron microscopy (SEM) and confocal optical micrographs of the resulting monolithic, free-standing CBGs are shown in \cref{fig:Structural_Characterization}a.

EBL is performed using a \qty{100}{\nm}-thick poly(methyl methacrylate) (PMMA) resist layer, employing an acceleration voltage of \qty{5}{\kilo\volt}, a \qty{30}{\micro\metre} aperture, and an exposure dose of \qty{100}{\micro\coulomb\per\centi\metre\squared}. Spatial alignment to the AFM-defined oxide markers ensures the deterministic placement of the cavity center relative to the buried QD \cite{Dhurjati2026}. Pattern transfer is executed via inductively coupled plasma reactive ion etching (ICP-RIE) using a $\mathrm{Cl}_2/\mathrm{Ar}$ chemistry (flow rates of \qty{2}{sccm} and \qty{18}{sccm}, respectively). The process is carried out at a chamber pressure of \qty{2}{\milli\bar}, an RF power of \qty{100}{\watt}, an ICP power of \qty{50}{\watt}, a DC self-bias of \qty{340}{\volt}, and a chuck temperature of \qty{0}{\celsius}. These conditions yield near-anisotropic trench profiles (see FIB cross-section in \cref{fig:Structural_Characterization}a) with etch rates of approximately \qty{5}{\nm\per\second} for \AlGaAs{}/\GaAs{} and \qty{1.5}{\nm\per\second} for PMMA, resulting in an effective mask selectivity of approximately \num{3.3}.

Membrane release is subsequently performed by immersing the sample in \qty{12.5}{\percent} hydrofluoric acid (HF) for \qty{20}{\second}. The HF rapidly enters through the deep under-etch holes, selectively removing the $\mathrm{Al}_{0.75}\mathrm{Ga}_{0.25}\mathrm{As}$ sacrificial layer below the CBG to yield a fully suspended membrane. A brief rinse in \qty{0.86}{\percent} potassium hydroxide (KOH) is utilized to dissolve residual inorganic aluminum fluoride by-products formed during the HF etch. Finally, the sample is rinsed in deionized water, transferred to isopropanol, and dried using supercritical $\mathrm{CO}_2$ critical point drying to prevent the suspended membranes from collapsing under capillary forces.

\subsection{Aspect-Ratio-Dependent Etching}

A fundamental requirement of our fabrication process is the simultaneous formation of shallow grating trenches (targeting roughly \qty{68}{\percent} of the membrane thickness to create vertical asymmetry) and deep under-etch openings (\textgreater\qty{150}{\nm} to expose the sacrificial layer) in a single step. To accomplish this, we exploit aspect-ratio-dependent etching (ARDE), also referred to as RIE lag. In this regime, narrow features etch more slowly than wider openings due to the transport limitations of reactive species and volatile etch by-products within high-aspect-ratio structures.

\begin{figure}[H]
    \centering
    \includegraphics[width=0.8\textwidth]{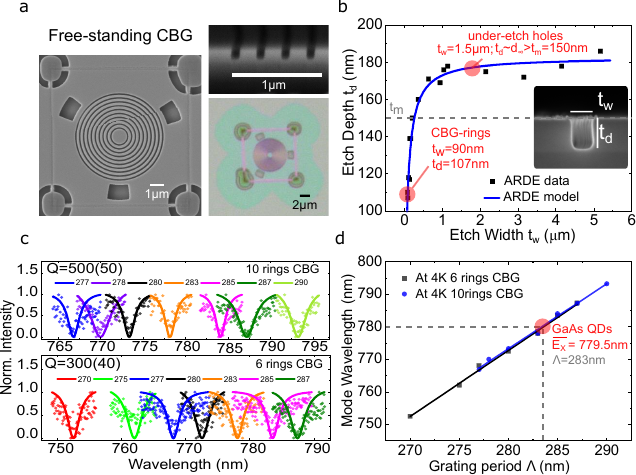}
    \caption{\label{fig:Structural_Characterization}
    Structural and optical characterization of circular Bragg gratings (CBGs) with deterministically positioned quantum dots (QDs).
    (a) Scanning electron microscopy (SEM) image of a QD integrated within a CBG and positioned relative to AFM-defined oxide markers, whose corners define the alignment reference frame. The top inset shows a focused ion beam (FIB) cross-section of the trench profile, while the bottom inset shows a confocal microscope image of a free-standing CBG, with the highlighted region indicating the suspended membrane.
    (b) Aspect ratio dependent etching (ARDE) characterized using an empirical model, enabling accurate prediction of trench depth as a function of feature width. Etch parameters are optimized to achieve target geometries for both CBG rings and under-etch holes. The inset shows a cleaved trench SEM image used for extraction of trench width and depth.
    (c–d) White-light reflectance spectroscopy of the CBG cavity modes as a function of grating period ($\Lambda$), measured for 6-ring and 10-ring devices at a temperature of \qty{4}{K}. The observed linear mode dispersion confirms controlled spectral tuning via $\Lambda$.
    }
\end{figure}

The ARDE behavior of the RIE process is characterized using SEM images of trenches with varying widths obtained from cleaved calibration samples. Trench widths $t_w$ and corresponding etch depths $t_d$ are extracted and analyzed to determine the functional interdependence. The experimental dataset shown in \cref{fig:Structural_Characterization}b is obtained by characterizing $t_d$ on a \AlGaAs membrane test sample subjected to the above-mentioned RIE-process for \qty{40}{\second}. The data is well described by the following transport-limited model, which captures the observed ARDE behavior \cite{Lai2006, Abraham1999, McNevin1998}:
\begin{equation}
\label{eq:ARDE_depth}
d(w) = d_{\infty} \, \frac{w}{w + w_{c}} \, .
\end{equation}

Here $d_{\infty}$ represents the saturation etch depth in the limit of large trench widths (i.e. unaffected by ARDE) and $w_{c}$ is the characteristic width at which the etch depth drops to $d_{\infty}/2$. By modeling the dataset to \cref{eq:ARDE_depth} the following parameters are obtained: \( d_{\infty} = \qty{183}{\nm} \) and \( w_{c} = \qty{57}{\nm} \) with a coefficient of determination is $R^2 = 0.96$.

This calibration establishes a quantitative link between the lithographic design and the realized 3D cavity geometry. For instance, the FDTD simulations for a \qty{150}{\nm}-thick membrane dictate an ideal CBG trench width of $t_W = \qty{83}{\nm}$ and a depth of $t_d = \qty{103}{\nm}$ (\cref{tab:CBG_parameters}). Inputting $t_W = \qty{83}{\nm}$ into \cref{eq:ARDE_depth} predicts an ARDE-limited etch depth of approximately \qty{108}{\nm}, closely matching the simulation target. Indeed, post-fabrication SEM measurements of the actual devices indicate realized dimensions of $t_W^\text{fab} = \qty{90(5)}{\nm}$ and $t_d^\text{fab} = \qty{100(5)}{\nm}$, confirming highly accurate pattern transfer within \qty{10}{\percent} of the idealized design. Concurrently, the wide outer under-etch openings (approximately \qty{1.5}{\micro\metre} wide) operate in the ARDE-free regime, reliably etching to $d_{\infty} = \qty{176}{\nm}$. This successfully breaches the \qty{150}{\nm} \AlGaAs{} membrane to expose the sacrificial layer below, validating the efficacy of the single-step ARDE-assisted geometry.

\subsection{Structural Characterization}

Near-infrared broadband reflectivity measurements are employed to characterize the optical properties of the fabricated CBG modes. Using a thermal light source (halogen lamp) focused onto the structures through a microscope objective, the reflected signal is spectrally resolved and normalized against unpatterned reference regions. This normalization procedure compensates for the wavelength-dependent system response, isolating the cavity resonances (\cref{fig:Structural_Characterization}d). 

Measurements were performed at a cryogenic temperature of \qty{4}{K} on reference CBGs fabricated without embedded QDs. By correlating the fundamental cavity resonance wavelength ($\lambda_{\mathrm{cav}}$) with the grating period ($\Lambda$), we calibrate the effective scaling of the photonic mode. As shown in \cref{fig:Structural_Characterization}c, $\lambda_{\mathrm{cav}}$ exhibits a highly predictable linear dependence on $\Lambda$, described by the linear model $\lambda_{\mathrm{cav}}(\Lambda) = a + b\,\Lambda$. Parameter estimations yields slopes and intercepts of $b = \num{2.03(5)}$ and $a = \num{205(15)}$ for \num{6}-ring devices, and $b = \num{1.99(6)}$ and $a = \num{215(18)}$ for \num{10}-ring devices, respectively. This empirical relation provides a direct design rule for tuning the grating period to match the emission wavelength of the \GaAs \, quantum dots. The average exciton energy of planar QDs is $\overline{E_X} = \qty{1.590(4)}{\eV}$, corresponding to $\qty{779.5}{\nm}$. Substituting this target wavelength into the linear model relation yields an calibrated target grating period of $\Lambda \to \qty{283}{\nm}$. An overview of the simulated and fabricated structural CBG parameters is provided in \cref{tab:CBG_parameters}.

Reflectance spectroscopy investigrations are performed for CBGs comprising both \num{6} and \num{10} grating rings. While \num{6}-ring devices exhibited Q-factors of \num{300(40)}, \num{10}-ring CBGs yielded Q-factors of approximately \num{500(50)}. To ensure robust optical confinement against potential fabrication-induced symmetry breaking or minor under-etch variations near the alignment markers, the more conservative \num{10}-ring design was employed for the final deterministically integrated QD devices.


\begin{table}[]
    \centering
    \sisetup{table-format=<3.0}
    \begin{tabular}{lcccc|cccc}
               & \multicolumn{4}{l}{Structural parameters}            & \multicolumn{4}{l}{Simulation results}                                               \\
               & $t_M$ (nm) & $t_W$ (nm) & $\Lambda$ (nm) & $t_d$ (nm) & $Q$       & $F_p$     & \EEfreeNom & \EEfiberNom \\
                \hline
    optimized  & \num{134}  & \num{77}   & \num{276}     & \num{83}   & \num{212} & \num{9.4} & \num{0.55} & \num{0.40}  \\
    optimized  & \num{150} & \num{83}   & \num{270}     & \num{103}  & \num{265} & \num{11.1}  & \num{0.50} & \num{0.31}   \\
    fabricated & \num{150(2)}  & \num{92(5)} & \num{283(2)} & \num{105(5)}  & \num{253} & \num{10.1}  & \num{0.43} & \num{0.27}                   
    \end{tabular}
    \label{tab:CBG_parameters}
    \caption{Overview of the CBG parameters and FTDT simulation results of optimized and fabricated devices, see \cref{sec:concept} and \cref{sec:fabrication}, respectively. $t_M$ is the \AlGaAs membrane thickness, $t_W$ the CBG trench width, $\Lambda$ the CBG grating periode, $t_d$ the CBG trench depth, $Q$ the cavity mode quality factor, $F_p$ the Purcell factor and EE the extraction efficiency. All simulated devices feature $N_r = \num{6}$ rings and target fundamental mode wavelengths of \qty{780}{\nm}. The fabricated devices use $N_r = \num{10}$ to compensate for possible manufacturing defects. The figure of merit for the optimizations is the extraction efficiency into a lensed fiber \EEfreeNom.}
    
\end{table}

\section{Post-Fabrication Device Performance}
\label{sec:Fab_Device_Perform}

To quantitatively assess the impact of fabrication tolerances on the optical performance of the realized devices, post-fabrication FDTD simulations are performed. These simulations incorporate the experimentally verified geometrical dimensions obtained from SEM analysis (\cref{tab:CBG_parameters}). In contrast to the ideal target parameters for a \qty{150}{\nm}-thick membrane, the experimentally realized structures exhibit slightly widened trench widths of $t_W = \qty{92(5)}{\nm}$ and corresponding trench depths of $t_d = \qty{105(5)}{\nm}$, alongside a measured membrane thickness of $t_M = \qty{150(2)}{\nm}$.

\begin{figure}[H]
    \centering
    \includegraphics[width=0.8\textwidth]{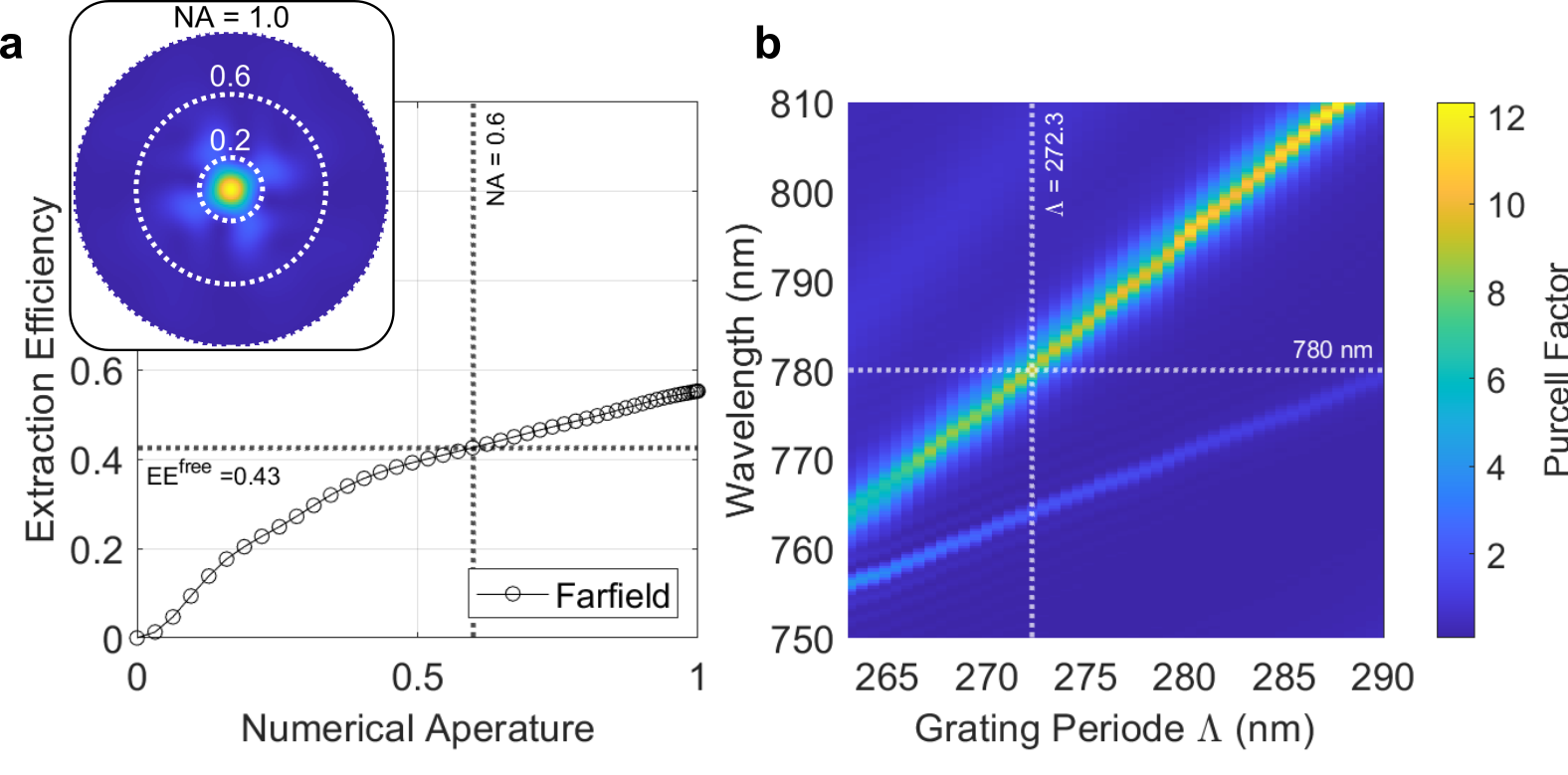}
    \caption{\label{fig:Fab_Sim}
    Simulated optical performance of the experimentally realized monolithic, free-standing CBG architecture. The device parameters reflect the post-fabrication measured dimensions (\cref{tab:CBG_parameters}). (a) Free-space extraction efficiency as a function of the collection numerical aperture (NA). The efficiency within an NA of \num{0.6} is \num{0.43}. Inset: Squared electrical field amplitude of the CBG far-field emission in polar representation. (b) Purcell factor of the centrally located dipole source as a function of emission wavelength and grating period $\Lambda$.}
\end{figure}

Simulations employing these measured structural dimensions predict a modest reduction in the extraction efficiency compared to the optimized ideal \qty{150}{\nm} design. Specifically, the simulated free-space collection efficiency (\EEfreeNom) drops from \qty{50}{\percent} to \qty{43}{\percent}, while the fiber-coupled efficiency (\EEfiberNom) is reduced from \qty{31}{\percent} to \qty{27}{\percent}. 

Furthermore, the FDTD models indicate that to achieve fundamental cavity mode resonance at the target QD emission wavelength of $\lambda = \qty{780}{\nm}$, a grating period of $\Lambda = \qty{272.3}{\nm}$ would be required. However, the experimental reflectivity calibrations (\cref{sec:fabrication}) demonstrate that a grating period of $\Lambda \approx \qty{283}{\nm}$ is actually required to reach this target wavelength. This spectral discrepancy between the idealized \textit{in silico} model and the experimental realization can be attributed to complex fabrication artifacts not captured by the idealized orthogonal FDTD geometry. These factors include non-ideal RIE etching anisotropy (such as slight sidewall inclination), localized variations in the membrane under-etch profile, surface roughness, and uncertainties in the true cryogenic refractive index of the processed \AlGaAs{} membrane. 

Despite these inevitable fabrication-induced deviations, the FDTD analysis confirms that the free-standing CBG architecture is remarkably robust. The fabricated devices retain highly directional vertical emission characteristics (\cref{fig:Fab_Sim}a) and broadband cavity quality factors suitable for simultaneous exciton-biexciton enhancement. Ultimately, an expected fiber-coupling efficiency of \qty{27}{\percent} from a fully monolithic, strain-free platform represents a highly competitive benchmark. The combination of deterministic AFM-NL positioning, a single-step RIE fabrication flow, and vertical asymmetry induced by partial etching establishes this platform as a highly pragmatic and scalable alternative to conventional mirror-bonded microcavities.

\section{Optical Characterisation}
\label{sec:opt_charac}

Optical characterization of the fabricated CBGs is performed using a micro-photoluminescence ($\mu$PL) setup at a cryogenic temperature of \qty{4}{K} (\cref{fig:Setup_Fabrication}). The devices are optically pumped using non-resonant, above-band excitation to probe the emission from the embedded QDs and to evaluate the cavity-mediated modifications to their optical properties. The QDs are excited to the saturation level of the neutral exciton ($X$) transition \cite{Hopfmann2021a} utilizing a \qty{635}{\nm} continuous-wave laser combined with a $90:10$ beam splitter. Excitation and luminescence collection are conducted through a near-infrared (NIR) microscope objective with a numerical aperture of \num{0.55}. The collected luminescence is filtered by a \qty{700}{\nm} long-pass filter and spectrally resolved using a \qty{0.75}{\metre} focal length spectrometer equipped with a \qty{1200}{lines/\mm} diffraction grating, achieving a spectral resolution of approximately \qty{35}{\micro\eV}.

\begin{figure}[H]
    \centering
    \includegraphics[width=0.7\textwidth]{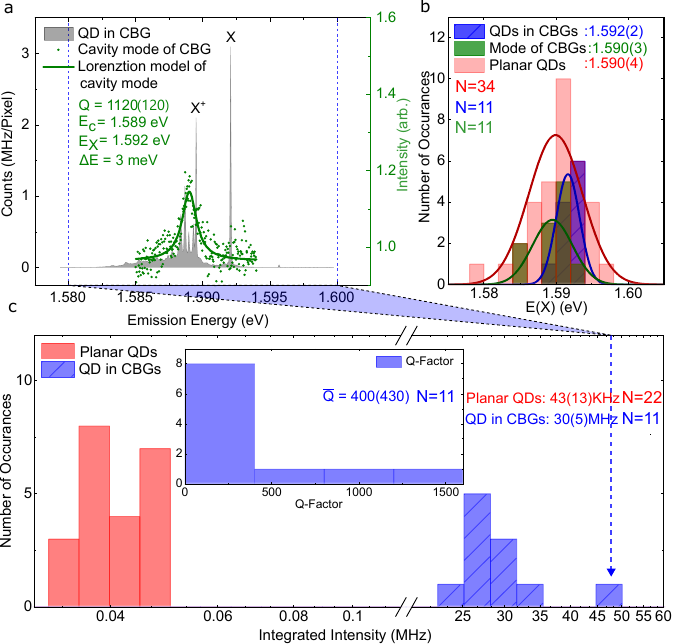}
    \caption{\label{fig:Optical_Characterization}
    Optical performance of circular Bragg gratings (CBGs) with deterministically positioned quantum dots (QDs).
    (a) $\mu$-photoluminescence ($\mu$-PL) spectrum of a representative bright device showing exciton (X) and positively charged trion (X$^+$) emission. The inverted reflection spectrum of the same representative CBG and its modeling to a Lorentzian curve is shown in green as an overlay. The spectral range employed for deriving the integrated intensity values is indicated by the blue dashed lines. (b) Statistical distribution of the X emission energies of QDs in CBGs, of CBG modes and planar reference QDs. (c) Statistical distribution of integrated emission intensities for planar QDs and CBG-coupled QDs from a single fabrication batch, demonstrating enhanced extraction efficiency and high device yield. The inset depicts the distribution of CBG Q-factors.
    }
\end{figure}

The $\mu$PL emission spectrum of a representative bright CBG device is presented in \cref{fig:Optical_Characterization}a. For this specific device, the $X$ emission line exhibits an intensity enhancement of roughly two to three orders of magnitude compared to typical QDs in unpatterned planar regions of the same sample. An overlaid Lorentzian cruve of the fundamental CBG cavity mode, extracted from the inverse reflectance spectrum, yields a cavity resonance energy of $E_C = \qty{1.589}{\eV}$ ($\lambda_{C} = \qty{780.3}{\nm}$) and a quality factor of $Q = \num{1120(120)}$. The QD exciton emission is observed at $E_X = \qty{1.592}{\eV}$ ($\lambda_{X} = \qty{778.8}{\nm}$), demonstrating excellent spectral alignment between the emitter and the broadband cavity mode.

To comprehensively benchmark the extraction performance, we evaluate the integrated emission intensity. This quantity is determined by numerically integrating the area of the $\mu$PL spectrum over a \qty{\pm 10}{\nm} spectral window centered around the $X$ transition. Applying this standardized procedure, which is fully analogous to our previous works \cite{Nie2021, Langer2025a}, an integrated intensity of \qty{45}{\mega\hertz} (detected counts per second) is obtained for the representative device shown in \cref{fig:Optical_Characterization}a. This analysis was extended to an ensemble of \num{11} fabricated CBGs and \num{22} reference planar QDs, with the resulting emission energies and integrated intensities summarized in \cref{fig:Optical_Characterization}b. The average exciton energy is $\overline{E_X} = \qty{1.590(4)}{\eV}$ for the planar QDs and $\qty{1.592(2)}{\eV}$ for QDs embedded in CBGs, while the average cavity mode energy is $\overline{E_C} = \qty{1.590(3)}{\eV}$. The strong overlap between $\overline{E_X}$ and $\overline{E_C}$ confirms the accuracy of the ARDE-based structural calibration. Due to the natural inhomogeneous broadening of the LDE growth process \cite{Keil2017} alongside subtle fabrication variations, only \num{6} out of the \num{11} devices exhibit optimal spectral overlap and strong emission enhancement. The fabricated ensemble yields an average cavity Q-factor of roughly \num{400} (inset, \cref{fig:Optical_Characterization}b). While this demonstrates excellent optical confinement, it is higher than the ideal target for broadband entangled photon pair sources, where lower Q-factors are preferred to simultaneously enhance both the $X$ and $XX$ transitions. Without active resonance tuning mechanisms, such as Stark effect or magnetic field tuning \cite{Ramanathan2013, Schnauber2021, Bayer1998}, this necessitates a reduction in the number of grating rings in future designs to further increase the spectral bandwidth and tolerance to emitter--cavity detuning.

The integrated intensities of QDs embedded in the CBGs exhibit massive enhancements over the unpatterned references (\cref{fig:Optical_Characterization}c). The planar QDs display an average integrated intensity of \qty{43(13)}{\kilo\hertz}, whereas the CBG-coupled QDs reach an average of \qty{30(5)}{\mega\hertz}, corresponding to an average intensity enhancement factor of roughly $\times \num{700}$. It is important to note that the reference planar QDs reside within a \qty{150}{\nm}-thick \AlGaAs{} membrane resting on an unetched sacrificial layer. This continuous membrane acts as a lateral 2D slab waveguide, trapping a significant fraction of the emitted light and reducing the vertical extraction efficiency below that of true bulk samples. While this structural artifact inflates the relative enhancement factor, the absolute performance metric---the maximum detected count rate of \qty{45}{\mega\hertz}---provides a definitive, application-relevant benchmark. This absolute brightness compares highly favorably to previously reported broadband solid-state quantum light sources, such as monolithic microlenses (\qty{30}{\mega\hertz}) and solid immersion lenses (\qty{27}{\mega\hertz}) \cite{Nie2021, Langer2025a}. These results confirm the highly competitive performance of the monolithic, free-standing CBG architecture.

\subsection{Fine Structure Splitting and Coherence Measurements}

\begin{figure}[H]
    \centering
    \includegraphics[width=\textwidth]{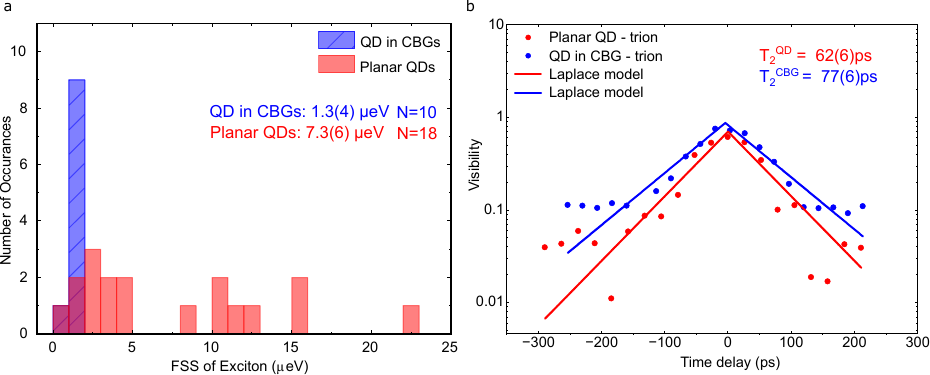}
    \caption{\label{fig:FSS_Coh}
    (a) Statistical distribution of exciton fine structure splitting (FSS) for planar QDs and QDs embedded in CBGs. (b) Exemprlary Michelson interferometer coherence time $T_2$ measurements and corresponding Laplacian visibility versus delay model curves of QD trion X$^+$ emission lines of a QD embedded in a CBG and a planar QD.}
\end{figure}

Beyond raw extraction efficiency, preserving the inherent optical qualities of the emitters--specifically the excitonic fine-structure splitting (FSS) and the photonic coherence time ($T_2$)--is critical to evaluating the suitability of the free-standing CBG devices for quantum networking. 

To evaluate the FSS, polarization-resolved spectroscopy is performed by introducing a motorized rotating half-wave plate and a fixed linear polarizer into the collection beam path. The FSS is determined by extracting the amplitude of the $X$ emission energy variation as a function of the waveplate angle $\alpha$. The resulting curve is normalized and modeled using a sinusoidal function: $\Delta E_{X}(\alpha) = A \sin(2\alpha+b)$, where the FSS is defined as $2A$. Further details on the modeling procedure and the associated Stokes parameter analysis are provided in the supplementary material \cref{sec:FSS}. As shown in \cref{fig:FSS_Coh}a, the planar reference QDs exhibit a significantly higher average FSS of \qty{7.3(61)}{\micro\eV}. In stark contrast, QDs embedded in the suspended CBGs display a heavily reduced average FSS of \qty{1.31(38)}{\micro\eV}, with all measured CBG devices yielding values below \qty{2.5}{\micro\eV}. This pronounced reduction strongly indicates that releasing the membrane into a free-standing architecture effectively relaxes residual epitaxial strain within the heterostructure. For entangled photon pair sources, such near-zero FSS values are highly advantageous, as they suppress exciton spin precession and are strictly essential for generating non-rotating entangled photon pair states suitable for high-fidelity entanglement swapping \cite{Winik2017, Hopfmann2021, Yang2022}.

Finally, coherence time ($T_2$) measurements are conducted using a Michelson interferometer placed in the spectrometer beam path. To avoid beating oscillations in the visibility curve caused by the FSS in neutral excitons, the first-order coherence is exclusively evaluated using the positively charged trion ($X^+$) emission line (\cref{fig:FSS_Coh}b). The visibility of the interference fringes as a function of optical time delay is modeled using a Laplace decay function. The extracted coherence time represents the combined influence of the radiative lifetime ($T_1$) and pure inhomogeneous dephasing ($T_2^*$), governed by the relation $1/T_2 = 1/(2T_1) + 1/T_2^*$. Characterization across the batch yielded an average $T_2$ of \qty{70(25)}{\pico\second} for QDs in CBGs, closely matching the \qty{56(20)}{\pico\second} average measured for planar QDs (\cref{sec:CohM}). This equivalence confirms that the invasive HF membrane release and plasma etching steps do not introduce additional non-radiative decay channels or fluctuating charge traps near the emitter. Given that the typical $T_1$ for such QD excitonic complexes is approximately \qty{200}{\pico\second} \cite{Keil2017, Hopfmann2021}, the measured coherence under non-resonant continuous-wave excitation is strictly dominated by inhomogeneous dephasing driven by local charge fluctuations, meaning $1/(2T_1) \ll 1/T_2^*$ \cite{Kuhlmann2013, Zhai2020}. Consequently, any Purcell-induced reduction in $T_1$ inside the cavity is heavily masked by the dominant $T_2^*$ dephasing, explaining the lack of a distinct Purcell signature in the continuous-wave $T_2$ measurements. Future time-resolved and strictly resonant pulsed excitation measurements will be required to decouple these dynamic effects and fully quantify the Purcell enhancement of this platform.

\section{Conclusion}

We demonstrate a fabrication-minimal approach for realizing monolithic and free-standing circular Bragg grating (CBG) cavities deterministically aligned to pre-selected GaAs quantum dots (QDs). By combining AFM-based lithographic positioning with a single-step top-down nanofabrication process exploiting aspect-ratio-dependent etching, suspended CBG structures are obtained without the need for bottom reflectors, flip-chip bonding, or multilayer processing.

Optical characterization, including broadband reflectance, $\mu$PL spectroscopy, and statistical device analysis, confirms robust vertical emission and consistent emitter–cavity coupling. Enhancement factors of up to \num{700} relative to unprocessed QDs are observed, with moderate cavity quality factors ranging from \num{400} to \num{1120}. The measured integrated QD photoluminescence (PL) intensity reaches up to \qty{45}{\MHz}, exceeding previously reported values for comparable systems based on monolithic microlenses and GaP solid immersion lenses \cite{Langer2025a, Nie2021}. While this metric provides a useful benchmark, a definitive assessment of device performance requires resonant pulsed excitation and photon-correlation measurements, which remain subjects of future work.

Polarization-resolved measurements reveal reduced exciton fine-structure splitting (FSS) values of \qty{1.3(4)}{\micro eV} for QDs embedded in CBGs, compared to average values of \qty{7(6)}{\micro eV} in planar heterostructures. Such reduced FSS is advantageous for entangled photon pair generation, as it suppresses excitonic precession dynamics \cite{Winik2017, Hopfmann2021, Yang2022}. These results indicate that by suspending the CGB structure significantly reduces the strain within the center of the CBG. Further investirgations are needed to investigate this effect in more detail.

Coherence measurements yield average coherence times of \qty{70(25)}{ps} for QDs in CBGs and \qty{56(20)}{ps} for planar QDs, indicating that the fabrication process preserves emitter coherence within experimental uncertainty.

Finite-difference time-domain simulations are employed to optimize the CBG design for efficient coupling into single-mode fibers and to benchmark fabricated structures using experimentally extracted geometrical parameters. This approach establishes a quantitative link between fabrication tolerances and optical performance. Although deviations from ideal design parameters reduce the achievable extraction efficiency, the fabricated devices maintain strong vertical emission and broadband cavity characteristics. The results emphasize the importance of precise control over trench depth, membrane thickness, and structural symmetry.

Overall, the presented platform provides a scalable route toward integrated quantum photonic devices. Future work will focus on improving fabrication precision, optimizing cavity bandwidths, and implementing resonant excitation schemes alongside photon-correlation measurements to fully evaluate the suitability of these structures for applications such as entangled photon pair sources.

\section{\label{acknowledgment} Acknowledgment}
We acknowledge Yana Vaynzof for valuable discussions and suggestions. We thank the clean room team of the Leibniz IFW Dresden, especially Ronny Engelhard and Sandra Nestler, for their efforts and expertise in processing of samples. This work was funded by the German federal ministry of research technology and space (BMFTR) projects QR.X, QUARKS, QUIET, and QD-CamNetz (contracts no. 16KISQ016, 16KIS1998K, 16KISQ094, and 16KISQ078).

\section*{\label{das} Data Availability Statement}
All data that support the findings presented in this work are available from the corresponding author upon reasonable request.

\section*{Conflict of Interest}
All authors declare that they have no conflicts of interest.

\section*{Supplementary}

\subsection{Polarisation-resolved Measurements}
\label{sec:FSS}

\begin{figure}[H]
    \centering
    \includegraphics[width=0.8\textwidth]{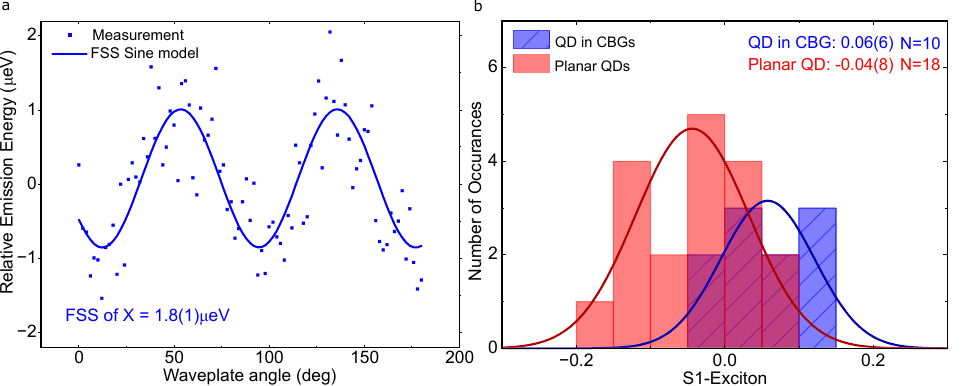}
    \caption{\label{fig:FSS}
    Excitonic fine structure splitting (FSS) and distribution of linear Stokes parameter ($S_1 = \frac{I_H - I_V}{I_H + I_V}$).
    (a) FSS of representative CBG modeled with sine function shown in the text giving a value of \qty{1.8(1)}{\mu eV}. (b) Distribution of $S_1$ of the exciton emission line for QDs in CBGs and planar QDs. }
\end{figure}

In order to evaluate the impact of the CBG devices on the observed QD fine-structure splitting (FSS) polarisation-resolved spectroscopy measurements are performed. This is achieved by the motorized rotation of a half-wave plate in front of a polarizer placed before the spectrometer, cf. \cref{fig:Setup_Fabrication}. The exciton FSS is determined by extracting the amplitude of the energetic variation of the QD X emission line as a function of the waveplate angle by modeling the data to $\Delta E(\alpha) = A \, sin(2\alpha+b)$. The FSS therefore corresponds to $\Delta E_\text{FSS} = 2A$. This type of measurement is exemplarily show in \cref{fig:FSS}a. By analysis of the Stokes parameters of the QD trionic emission the lateral emitter–cavity misalignment and its impact on polarization and entanglement properties can be investigated \cite{Buchinger2025, Dhurjati2026}. Due to the CBG circular geometry \cite{Kavokin2007}, the linear Stokes parameter $S_1 = \frac{I_H - I_V}{I_H + I_V}$ is the figure of merit for benchmarking this effect. The statistical destributions of $S_1$ of QD in CBGs fabricated for this study and of reference planar QDs are shown in \cref{fig:FSS}b. For QDs embedded in CBGs, the Stokes parameter $S_1$ exhibits a mean value of \num{0.06(6)}, while planar QDs yield a value of \num{-0.04(8)}. The near-zero values in both cases indicate minimal polarization imbalance, consistent with good spatial alignment of the emitters within the cavity. A detailed analysis of positioning accuracy of AFM Nanolithography technique and its influence on polarization properties is presented in our recent work Ref. \cite{Dhurjati2026}. 

\subsection{Coherence Time Measurements}
\label{sec:CohM}

\begin{figure}[H]
    \centering
    \includegraphics[width=0.8\textwidth]{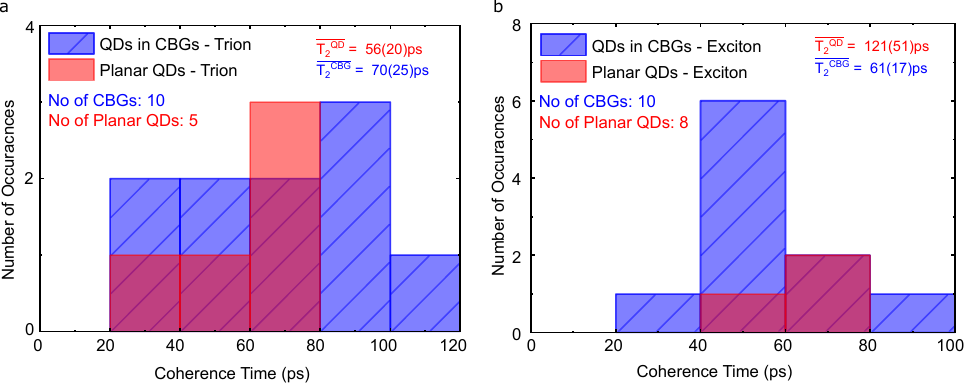}
    \caption{\label{fig:T2_stat}
    Coherence measurements of QD in CBGs.
    (a)Distribution of measured coherence time of planar QDs and CBGs for of trion line. (b)  Distribution of measured coherence time of planar QDs and CBGs for of exciton line.}
\end{figure}

Coherence measurements were performed using interferometric visibility analysis using a Michelson interferometer in front of the spectrometer, cf. \cref{fig:Setup_Fabrication}. The statistical results for both QD $X$ and $X^+$ emission lines and for QDs in CBGs and planar hetrostructures are summarized in \cref{fig:T2_stat}. The first-order coherence of the emission is extracted from the visibility of interference fringes as a function of time delay, with the data modeled using a Laplace decay model for both $X$ and $X^+$ transitions, as shown in \cref{fig:FSS_Coh}b. The extracted coherence times reflect the combined influence of radiative lifetime and dephasing processes within the emitter--cavity system.

The measured coherence times yield average values of \qty{56(20)}{\pico\second} for planar QDs and \qty{70(25)}{\pico\second} for CBG-coupled QDs. The comparable coherence properties of both processed and unprocessed QDs indicates that integration into the CBG does not introduce additional dephasing mechanisms. On the contrary, due to the lifetime ($T_1$) reduction by the Purcell effect in microcavities one would expect a reduction of the $T_2$ values in the case of the QDs in CBGs using the relation $T_2 = (\frac{1}{2 T_1}+\frac{1}{T_2^*})^{-1}$. As a consequence of this observation one can conclude that inhomogenious dephasing $T_2^*$, i.e. due to charge noise, limits the coherence of both QDs in CBGs and planar structures, i.e. $T_2^* \ll 2T_1$. The observed $T_2$ values therefore indicate that the fabrication process does not significantly influence $T_2^*$, this means that there is no indication for manufacturing-induced addional depahsing. To evaluate the influence of the Purcell effect of $T_2$ additional investigations using (resonant) pump-probe lifetime measurments beyond the scope of this work are required.  

\subsection{Optimiaztion of CBG Devices with \qty{150}{\nm} membrane}
\label{sec:150nmSims}

 \begin{figure}[H]
    \centering
    \includegraphics[width=0.8\textwidth]{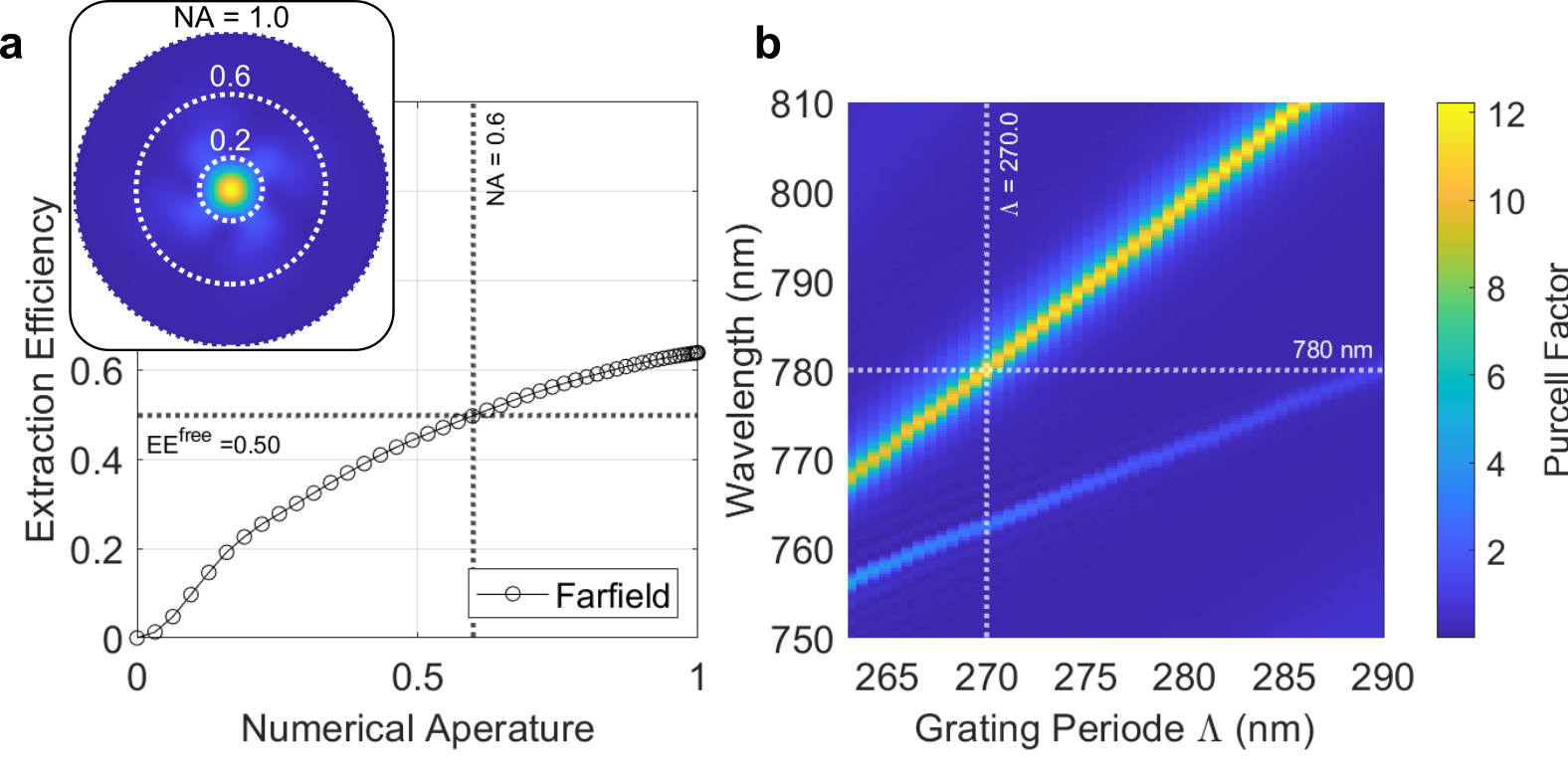}
    \caption{\label{fig:Ideal_150nm_FDTD Simulations}
    Simulated device performance of an optimzed monolithic, free-standing CBG featuring a \qty{150}{\nm} thick membrane, the device parameters are listed in \cref{tab:CBG_parameters}. The optimization figure of merit is \EEfiber, details are given in the text. (a) Free-space extraction efficiency as a function of the emission angle in terms of the numerical aperture (NA), the maximal obtained value at an NA of \num{0.6} is \num{0.50}. Inset: Squared electrical field amplitude of the CBG farfied emission in polar representation. (b) Purcell factor of the dipole source versus its emission wavelength and grating period $\Lambda$.}
\end{figure}

Due to technical limitations MBE grown substrates with the ideal thickness of \qty{134}{\nm} are unavailable. There are however heterostructures available with \qty{150}{\nm} membrane thickness. In order to attain the best possible performance using the available structures, the strutural parameters are optimized using finite-difference time-domain (FDTD) simulations to obtain the maximal fiber-coupled emission efficiency (\EEfiber) using a lensed fibers featuring an numerical aperature (NA) of \num{0.6}. The resuling set of parameters are listed in \cref{tab:CBG_parameters}, while the simulation results are visulaized in \cref{tab:CBG_parameters}. The obtainable \EEfiber is \qty{31}{\percent}, which is \qty{9}{\percent} lower than the \qty{134}{\nm} thick membranes.

\printbibliography

@Article{Claudon2010,
  author       = {Claudon, Julien and Bleuse, Joël and Malik, Nitin Singh and Bazin, Maela and Jaffrennou, Périne and Gregersen, Niels and Sauvan, Christophe and Lalanne, Philippe and Gérard, Jean-Michel},
  date         = {2010},
  journaltitle = {Nature Photonics},
  title        = {A highly efficient single-photon source based on a quantum dot in a photonic nanowire},
  doi          = {10.1038/nphoton.2009.287x},
  issn         = {1749-4893},
  number       = {3},
  pages        = {174--177},
  url          = {https://doi.org/10.1038/nphoton.2009.287x},
  volume       = {4},
  journal      = {Nature Photonics},
  year         = {2010},
}

@Article{Dousse2010,
  author       = {Dousse, Adrien and Suffczyński, Jan and Beveratos, Alexios and Krebs, Olivier and Lemaître, Aristide and Sagnes, Isabelle and Bloch, Jacqueline and Voisin, Paul and Senellart, Pascale},
  date         = {2010},
  journaltitle = {Nature},
  title        = {Ultrabright source of entangled photon pairs},
  doi          = {10.1038/nature09148},
  issn         = {1476-4687},
  number       = {7303},
  pages        = {217--220},
  url          = {https://doi.org/10.1038/nature09148},
  volume       = {466},
  journal      = {Nature},
  year         = {2010},
}

@Article{Somaschi2016,
  author       = {Somaschi, N. and Giesz, V. and De Santis, L. and Loredo, J. C. and Almeida, M. P. and Hornecker, G. and Portalupi, S. L. and Grange, T. and Antón, C. and Demory, J. and Gómez, C. and Sagnes, I. and Lanzillotti-Kimura, N. D. and Lemaítre, A. and Auffeves, A. and White, A. G. and Lanco, L. and Senellart, P.},
  date         = {2016},
  journaltitle = {Nature Photonics},
  title        = {Near-optimal single-photon sources in the solid state},
  doi          = {10.1038/nphoton.2016.23},
  issn         = {1749-4893},
  number       = {5},
  pages        = {340--345},
  url          = {https://doi.org/10.1038/nphoton.2016.23},
  volume       = {10},
  journal      = {Nature Photonics},
  year         = {2016},
}

@Article{Wang2019,
  author    = {Wang, Hui and Hu, Hai and Chung, T.-H. and Qin, Jian and Yang, Xiaoxia and Li, J.-P. and Liu, R.-Z. and Zhong, H.-S. and He, Y.-M. and Ding, Xing and Deng, Y.-H. and Dai, Qing and Huo, Y.-H. and H\"ofling, Sven and Lu, Chao-Yang and Pan, Jian-Wei},
  title     = {On-Demand Semiconductor Source of Entangled Photons Which Simultaneously Has High Fidelity, Efficiency, and Indistinguishability},
  doi       = {10.1103/PhysRevLett.122.113602},
  issue     = {11},
  pages     = {113602},
  url       = {https://link.aps.org/doi/10.1103/PhysRevLett.122.113602},
  volume    = {122},
  journal   = {Phys. Rev. Lett.},
  month     = {03},
  numpages  = {6},
  publisher = {American Physical Society},
  year      = {2019},
}

@Article{Liu2019,
  author   = {Liu, Jin and Su, Rongbin and Wei, Yuming and Yao, Beimeng and Silva, Saimon Filipe Covre da and Yu, Ying and Iles-Smith, Jake and Srinivasan, Kartik and Rastelli, Armando and Li, Juntao and Wang, Xuehua},
  title    = {A solid-state source of strongly entangled photon pairs with high brightness and indistinguishability},
  doi      = {10.1038/s41565-019-0435-9},
  issn     = {1748-3395},
  number   = {6},
  pages    = {586--593},
  url      = {https://doi.org/10.1038/s41565-019-0435-9},
  volume   = {14},
  journal  = {Nature Nanotechnology},
  year     = {2019},
}

@Article{Tomm2021,
  author       = {Tomm, Natasha and Javadi, Alisa and Antoniadis, Nadia Olympia and Najer, Daniel and Löbl, Matthias Christian and Korsch, Alexander Rolf and Schott, Rüdiger and Valentin, Sascha René and Wieck, Andreas Dirk and Ludwig, Arne and Warburton, Richard John},
  date         = {2021},
  journaltitle = {Nature Nanotechnology},
  title        = {A bright and fast source of coherent single photons},
  doi          = {10.1038/s41565-020-00831-x},
  issn         = {1748-3395},
  number       = {4},
  pages        = {399--403},
  url          = {https://doi.org/10.1038/s41565-020-00831-x},
  volume       = {16},
  journal      = {Nature Nanotechnology},
  year         = {2021},
}

@Article{Hopfmann2021,
  author    = {Hopfmann, Caspar and Nie, Weijie and Sharma, Nand Lal and Weigelt, Carmen and Ding, Fei and Schmidt, Oliver G.},
  title     = {Maximally entangled and gigahertz-clocked on-demand photon pair source},
  doi       = {10.1103/PhysRevB.103.075413},
  issue     = {7},
  pages     = {075413},
  url       = {https://link.aps.org/doi/10.1103/PhysRevB.103.075413},
  volume    = {103},
  journal   = {Phys. Rev. B},
  month     = {02},
  numpages  = {7},
  publisher = {American Physical Society},
  year      = {2021},
}

@Article{Huber2017,
  author    = {Huber, Daniel and Reindl, Marcus and Huo, Yongheng and Huang, Huiying and Wildmann, Johannes S and Schmidt, Oliver G and Rastelli, Armando and Trotta, Rinaldo},
  title     = {Highly indistinguishable and strongly entangled photons from symmetric {G}a{A}s quantum dots},
  number    = {1},
  pages     = {1--7},
  volume    = {8},
  journal   = {Nat. Commun.},
  publisher = {Nature Publishing Group},
  year      = {2017},
}

@article{Langer2025,
  author  = {Langer, M. and Ruchka, P. and Rahimi, A. and Jakovljevic, S. and Zena, Y. G. and Dhurjati, S. A. and Danilov, A. and Pal, M. and Bassoli, R. and Fitzek, F. H. P. and Schmidt, O. G. and Giessen, H. and Hopfmann, C.},
  title   = {An ultra-compact deterministic source of maximally entangled photon pairs},
  journal = {APL Photonics},
  volume  = {10},
  number  = {6},
  pages   = {066117},
  year    = {2025},
  doi     = {10.1063/5.0271023},
  url     = {https://doi.org/10.1063/5.0271023},
}

@Article{Langer2025a,
  author       = {Langer, Moritz and Dhurjati, Sai A and Zena, Yared G and Rahimi, Ahmad and Pal, Mandira and Raith, Liesa and Nestler, Sandra and Bassoli, Riccardo and Fitzek, Frank H P and Schmidt, Oliver G and Hopfmann, Caspar},
  date         = {2025-05},
  journaltitle = {Nanotechnology},
  title        = {Bright quantum dot light sources using monolithic microlenses on gold back-reflectors},
  doi          = {10.1088/1361-6528/add350},
  number       = {22},
  pages        = {225301},
  url          = {https://dx.doi.org/10.1088/1361-6528/add350},
  volume       = {36},
  publisher    = {IOP Publishing},
}

@article{Langer2025b,
  author        = {Moritz Langer and Sai Abhishikth Dhurjati and Martin Bauer and Yared Getahun Zena and Ahmad Rahimi and Riccardo Bassoli and Frank H. P. Fitzek and Oliver G. Schmidt and Caspar Hopfmann},
  title         = {Temperature-dependent refractive index of AlGaAs for quantum-photonic devices near the bandgap},
  eprint        = {2512.02212},
  url           = {https://arxiv.org/abs/2512.02212},
  archiveprefix = {arXiv},
  primaryclass  = {physics.optics},
  year          = {2025},
}

@Article{Bremer2020,
  author       = {Bremer, Lucas and Weber, Ksenia and Fischbach, Sarah and Thiele, Simon and Schmidt, Marco and Kaganskiy, Arsenty and Rodt, Sven and Herkommer, Alois and Sartison, Marc and Portalupi, Simone Luca and Michler, Peter and Giessen, Harald and Reitzenstein, Stephan},
  date         = {2020-10},
  journaltitle = {APL Photonics},
  title        = {{Quantum dot single-photon emission coupled into single-mode fibers with 3D printed micro-objectives}},
  doi          = {10.1063/5.0014921},
  issn         = {2378-0967},
  number       = {10},
  pages        = {106101},
  url          = {https://doi.org/10.1063/5.0014921},
  volume       = {5},
  journal      = {APL Photonics},
  year         = {2020},
}

@Article{Schlehahn2018,
  author       = {Schlehahn, Alexander and Fischbach, Sarah and Schmidt, Ronny and Kaganskiy, Arsenty and Strittmatter, André and Rodt, Sven and Heindel, Tobias and Reitzenstein, Stephan},
  date         = {2018},
  journaltitle = {Scientific Reports},
  title        = {A stand-alone fiber-coupled single-photon source},
  doi          = {10.1038/s41598-017-19049-4},
  issn         = {2045-2322},
  number       = {1},
  pages        = {1340},
  url          = {https://doi.org/10.1038/s41598-017-19049-4},
  volume       = {8},
  journal      = {Scientific Reports},
  year         = {2018},
}

@Article{Musial2020,
  author       = {Musiał, Anna and Żołnacz, Kinga and Srocka, Nicole and Kravets, Oleh and Große, Jan and Olszewski, Jacek and Poturaj, Krzysztof and Wójcik, Grzegorz and Mergo, Paweł and Dybka, Kamil and Dyrkacz, Mariusz and Dłubek, Michał and Lauritsen, Kristian and Bülter, Andreas and Schneider, Philipp-Immanuel and Zschiedrich, Lin and Burger, Sven and Rodt, Sven and Urbańczyk, Wacław and Sęk, Grzegorz and Reitzenstein, Stephan},
  date         = {2020},
  journaltitle = {Advanced Quantum Technologies},
  title        = {Plug\&Play Fiber-Coupled 73 {kHz} Single-Photon Source Operating in the Telecom O-Band},
  doi          = {https://doi.org/10.1002/qute.202000018},
  number       = {6},
  pages        = {2000018},
  url          = {https://onlinelibrary.wiley.com/doi/abs/10.1002/qute.202000018},
  volume       = {3},
  journal      = {Advanced Quantum Technologies},
  year         = {2020},
}

@Article{MoczalaDusanowska2020,
  author       = {Moczała-Dusanowska, Magdalena and Dusanowski, Łukasz and Iff, Oliver and Huber, Tobias and Kuhn, Silke and Czyszanowski, Tomasz and Schneider, Christian and Höfling, Sven},
  date         = {2020-12},
  journaltitle = {ACS Photonics},
  title        = {Strain-Tunable Single-Photon Source Based on a Circular Bragg Grating Cavity with Embedded Quantum Dots},
  doi          = {10.1021/acsphotonics.0c01465},
  number       = {12},
  pages        = {3474--3480},
  url          = {https://doi.org/10.1021/acsphotonics.0c01465},
  volume       = {7},
  publisher    = {American Chemical Society},
}

@Article{Holewa2024,
  author       = {Paweł Holewa and Daniel A. Vajner and Emilia Zięba-Ostój and Maja Wasiluk and Benedek Gaál and Aurimas Sakanas and Marek G. Mikulicz and Paweł Mrowiński and Bartosz Krajnik and Meng Xiong and Kresten Yvind and Niels Gregersen and Anna Musiał and Alexander Huck and Tobias Heindel and Marcin Syperek and Elizaveta Semenova},
  date         = {2024-04-18},
  journaltitle = {Nat. Commun.},
  title        = {High-throughput quantum photonic devices emitting indistinguishable photons in the telecom C-band},
  doi          = {10.1038/s41467-024-47551-7},
  number       = {1},
  pages        = {3358},
  url          = {https://doi.org/10.1038/s41467-024-47551-7},
  volume       = {15},
  publisher    = {Springer Nature},
}

@Article{Buchinger2025,
  author    = {Buchinger, Quirin and Krause, Constantin and Zhang, Aileen and Peniakov, Giora and Helal, Mohamed and Reum, Yorick and Pfenning, Andreas Theo and Höfling, Sven and Huber-Loyola, Tobias},
  title     = {Deterministic quantum dot cavity placement using hyperspectral imaging with high spatial accuracy and precision},
  journal   = {Nano Convergence},
  year      = {2025},
  volume    = {12},
  number    = {1},
  pages     = {36},
  doi       = {10.1186/s40580-025-00501-5},
  issn      = {2196-5404},
  url       = {https://doi.org/10.1186/s40580-025-00501-5},
}

@article{Sanguinetti2003,
  title   = {Modified droplet epitaxy GaAs/AlGaAs quantum dots grown on a variable thickness wetting layer},
  journal = {Journal of Crystal Growth},
  volume  = {253},
  number  = {1},
  pages   = {71-76},
  year    = {2003},
  doi     = {https://doi.org/10.1016/S0022-0248(03)01016-9},
  url     = {https://www.sciencedirect.com/science/article/pii/S0022024803010169},
  author  = {S. Sanguinetti and K. Watanabe and T. Tateno and M. Gurioli and P. Werner and M. Wakaki and N. Koguchi},
}

@article{Heyn2007,
  author  = {Heyn, Ch. and Stemmann, A. and Schramm, A. and Welsch, H. and Hansen, W. and Nemcsics, Á.},
  title   = {Faceting during GaAs quantum dot self-assembly by droplet epitaxy},
  journal = {Applied Physics Letters},
  volume  = {90},
  number  = {20},
  pages   = {203105},
  year    = {2007},
  doi     = {10.1063/1.2737123},
  url     = {https://doi.org/10.1063/1.2737123},
}

@article{Cambel2008,
  title   = {Local anodic oxidation by AFM tip developed for novel semiconductor nanodevices},
  journal = {Ultramicroscopy},
  volume  = {108},
  number  = {10},
  pages   = {1021-1024},
  year    = {2008},
  doi     = {https://doi.org/10.1016/j.ultramic.2008.04.032},
  url     = {https://www.sciencedirect.com/science/article/pii/S0304399108000697},
  author  = {Vladimír Cambel and Jozef Martaus and Ján Šoltýs and Robert Kúdela and Dagmar Gregušová},
}

@article{Cambel2007,
  author  = {Cambel, Vladimír and Šoltýs, Ján},
  title   = {The influence of sample conductivity on local anodic oxidation by the tip of atomic force microscope},
  journal = {Journal of Applied Physics},
  volume  = {102},
  number  = {7},
  pages   = {074315},
  year    = {2007},
  doi     = {10.1063/1.2794374},
  url     = {https://doi.org/10.1063/1.2794374},
}

@Article{Lodahl2015,
  author    = {Lodahl, Peter and Mahmoodian, Sahand and Stobbe, Søren},
  title     = {Interfacing single photons and single quantum dots with photonic nanostructures},
  doi       = {10.1103/RevModPhys.87.347},
  issue     = {2},
  pages     = {347--400},
  url       = {https://link.aps.org/doi/10.1103/RevModPhys.87.347},
  volume    = {87},
  journal   = {Rev. Mod. Phys.},
  publisher = {American Physical Society},
  year      = {2015},
}

@Article{Barbiero2022,
  author       = {Barbiero, Andrea and Huwer, Jan and Skiba-Szymanska, Joanna and Ellis, David J. P. and Stevenson, R. Mark and Müller, Tina and Shooter, Ginny and Goff, Lucy E. and Ritchie, David A. and Shields, Andrew J.},
  date         = {2022-09},
  journaltitle = {ACS Photonics},
  title        = {High-Performance Single-Photon Sources at Telecom Wavelength Based on Broadband Hybrid Circular Bragg Gratings},
  doi          = {10.1021/acsphotonics.2c00810},
  number       = {9},
  pages        = {3060--3066},
  url          = {https://doi.org/10.1021/acsphotonics.2c00810},
  volume       = {9},
  publisher    = {American Chemical Society},
}

@article{Ates2012,
  author    = {Ates, S. and Sapienza, L. and Davanco, M. and Badolato, A. and Srinivasan, K.},
  title     = {Bright Single-Photon Emission From a Quantum Dot in a Circular Bragg Grating Microcavity},
  journal   = {IEEE J. Sel. Top. Quantum Electron.},
  year      = {2012},
  volume    = {18},
  number    = {6},
  pages     = {1711--1721},
  doi       = {10.1109/JSTQE.2012.2193877},
  issn      = {1558-4542},
  url       = {https://doi.org/10.1109/JSTQE.2012.2193877},
  publisher = {IEEE},
}

@Article{Andreani1999,
  author    = {Andreani, Lucio Claudio and Panzarini, Giovanna and Gérard, Jean-Michel},
  title     = {Strong-coupling regime for quantum boxes in pillar microcavities: Theory},
  doi       = {10.1103/PhysRevB.60.13276},
  issue     = {19},
  pages     = {13276--13279},
  url       = {https://link.aps.org/doi/10.1103/PhysRevB.60.13276},
  volume    = {60},
  journal   = {Phys. Rev. B},
  publisher = {American Physical Society},
  year      = {1999},
}

@Article{Hopfmann2021a,
  author    = {Hopfmann, Caspar and Sharma, Nand Lal and Nie, Weijie and Keil, Robert and Ding, Fei and Schmidt, Oliver G.},
  title     = {Heralded preparation of spin qubits in droplet-etched {GaAs} quantum dots using quasiresonant excitation},
  doi       = {10.1103/PhysRevB.104.075301},
  issue     = {7},
  pages     = {075301},
  url       = {https://link.aps.org/doi/10.1103/PhysRevB.104.075301},
  volume    = {104},
  journal   = {Phys. Rev. B},
  publisher = {American Physical Society},
  year      = {2021},
}

@Article{Winik2017,
  author    = {Winik, R. and Cogan, D. and Don, Y. and Schwartz, I. and Gantz, L. and Schmidgall, E. R. and Livneh, N. and Rapaport, R. and Buks, E. and Gershoni, D.},
  title     = {On-demand source of maximally entangled photon pairs using the biexciton-exciton radiative cascade},
  doi       = {10.1103/PhysRevB.95.235435},
  issue     = {23},
  pages     = {235435},
  url       = {https://link.aps.org/doi/10.1103/PhysRevB.95.235435},
  volume    = {95},
  journal   = {Phys. Rev. B},
  publisher = {American Physical Society},
  year      = {2017},
}

@Article{Nie2021,
  author       = {Nie, Weijie and Sharma, Nand Lal and Weigelt, Carmen and Keil, Robert and Yang, Jingzhong and Ding, Fei and Hopfmann, Caspar and Schmidt, Oliver G.},
  date         = {2021},
  journaltitle = {Applied Physics Letters},
  title        = {Experimental optimization of the fiber coupling efficiency of {GaAs} quantum dot-based photon sources},
  doi          = {10.1063/5.0059310},
  number       = {24},
  pages        = {244003},
  url          = {https://doi.org/10.1063/5.0059310},
  volume       = {119},
  journal      = {Applied Physics Letters},
  year         = {2021},
}

@Article{Rickert2025,
  author       = {Rickert, Lucas and Żołnacz, Kinga and Vajner, Daniel A. and von Helversen, Martin and Rodt, Sven and Reitzenstein, Stephan and Liu, Hanqing and Li, Shulun and Ni, Haiqiao and Wyborski, Paweł and Sęk, Grzegorz and Musiał, Anna and Niu, Zhichuan and Heindel, Tobias},
  date         = {2025-01},
  journaltitle = {Nanophotonics},
  title        = {A fiber-pigtailed quantum dot device generating indistinguishable photons at GHz clock-rates},
  doi          = {10.1515/nanoph-2024-0519},
  issn         = {2192-8614},
  url          = {http://dx.doi.org/10.1515/nanoph-2024-0519},
  publisher    = {Walter de Gruyter GmbH},
}

@Article{Davanco2011,
  author   = {Davanço, M. and Rakher, M. T. and Schuh, D. and Badolato, A. and Srinivasan, K.},
  title    = {A circular dielectric grating for vertical extraction of single quantum dot emission},
  doi      = {10.1063/1.3615051},
  issn     = {0003-6951},
  number   = {4},
  pages    = {041102},
  url      = {https://doi.org/10.1063/1.3615051},
  volume   = {99},
  journal  = {Applied Physics Letters},
  month    = {07},
  year     = {2011},
}

@Misc{Zena2026,
  author      = {Yared G. Zena and Moritz Langer and Ahmad Rahimi and Sai Abhishikth Dhurjati and Pavel Ruchka and Sara Jakovljevic and Mandira Pal and Frank H. P. Fitzek and Harald Giessen and Juergen Czarske and Riccardo Bassoli and Caspar Hopfmann},
  date        = {2026},
  title       = {Compact system development of efficient quantum-entangled photon sources towards deployable and industrial devices},
  eprint      = {2604.02024},
  eprintclass = {quant-ph},
  eprinttype  = {arXiv},
  url         = {https://arxiv.org/abs/2604.02024},
}

@Misc{Dhurjati2026,
  author = {Sai Abhishikth Dhurjati and Moritz Langer Yared G. Zena and Ahmad Rahimi and Liesa Raith and Martin Bauer and Yana Vaynzof and Frank H. P. Fitzek and Riccardo Bassoli and Caspar Hopfmann},
  date   = {2026},
  title  = {Deterministic positioning of circular Bragg gratings using atomic force lithography for high-performance quantum dot light sources},
}

@Article{Yang2022,
  author       = {Yang, Jingzhong and Zopf, Michael and Li, Pengji and Sharma, Nand Lal and Nie, Weijie and Benthin, Frederik and Fandrich, Tom and Rugeramigabo, Eddy P. and Hopfmann, Caspar and Keil, Robert and Schmidt, Oliver G. and Ding, Fei},
  date         = {2022-06},
  journaltitle = {Phys. Rev. B},
  title        = {Statistical limits for entanglement swapping with semiconductor entangled photon sources},
  doi          = {10.1103/PhysRevB.105.235305},
  issue        = {23},
  pages        = {235305},
  url          = {https://link.aps.org/doi/10.1103/PhysRevB.105.235305},
  volume       = {105},
  numpages     = {9},
  publisher    = {American Physical Society},
}

@Article{Keil2017,
  author       = {Keil, Robert and Zopf, Michael and Chen, Yan and H{\"o}fer, Bianca and Zhang, Jiaxiang and Ding, Fei and Schmidt, Oliver G},
  date         = {2017},
  journaltitle = {Nat. Commun.},
  title        = {Solid-state ensemble of highly entangled photon sources at rubidium atomic transitions},
  number       = {1},
  pages        = {1--8},
  volume       = {8},
  publisher    = {Nature Publishing Group},
}

@Book{Michler2014,
  author    = {Michler, P.},
  date      = {2014},
  title     = {Single Quantum Dots: Fundamentals, Applications and New Concepts},
  isbn      = {9783662307861},
  publisher = {Springer Berlin Heidelberg},
  series    = {Topics in Applied Physics},
  url       = {https://books.google.de/books?id=5P0QswEACAAJ},
}

@Book{Kavokin2007,
  author    = {Alexey Kavokin and Jeremy J. Baumberg and Guillaume Malpuech and Fabrice P. Laussy},
  date      = {2007},
  title     = {Microcavities},
  doi       = {10.1093/acprof:oso/9780199228942.001.0001},
  publisher = {Oxford University Press},
}

@Article{Zhai2020,
  author       = {Zhai, Liang and Löbl, Matthias C. and Nguyen, Giang N. and Ritzmann, Julian and Javadi, Alisa and Spinnler, Clemens and Wieck, Andreas D. and Ludwig, Arne and Warburton, Richard J.},
  title        = {Low-noise GaAs quantum dots for quantum photonics},
  issn         = {2041-1723},
  number       = {1},
  pages        = {4745},
  volume       = {11},
  date         = {2020},
  doi          = {10.1038/s41467-020-18625-z},
  journaltitle = {Nature Communications},
  refid        = {Zhai2020},
  url          = {https://doi.org/10.1038/s41467-020-18625-z},
}

@Article{Rickert2023,
  author       = {Lucas Rickert and Fridtjof Betz and Matthias Plock and Sven Burger and Tobias Heindel},
  date         = {2023-04},
  journaltitle = {Opt. Express},
  title        = {High-performance designs for fiber-pigtailed quantum-light sources based on quantum dots in electrically-controlled circular Bragg gratings},
  doi          = {10.1364/OE.486060},
  number       = {9},
  pages        = {14750--14770},
  url          = {https://opg.optica.org/oe/abstract.cfm?URI=oe-31-9-14750},
  volume       = {31},
  publisher    = {Optica Publishing Group},
}

@Article{Schnauber2021,
  author       = {Schnauber, Peter and Große, Jan and Kaganskiy, Arsenty and Ott, Maximilian and Anikin, Pavel and Schmidt, Ronny and Rodt, Sven and Reitzenstein, Stephan},
  date         = {2021-05},
  journaltitle = {APL Photonics},
  title        = {Spectral control of deterministically fabricated quantum dot waveguide systems using the quantum confined Stark effect},
  doi          = {10.1063/5.0050152},
  issn         = {2378-0967},
  number       = {5},
  pages        = {050801},
  url          = {https://doi.org/10.1063/5.0050152},
  volume       = {6},
}

@Article{Ramanathan2013,
  author       = {Ramanathan, Swati and Petersen, Greg and Wijesundara, Kushal and Thota, Ramana and Stinaff, E. A. and Kerfoot, Mark L. and Scheibner, Michael and Bracker, Allan S. and Gammon, D.},
  date         = {2013-05},
  journaltitle = {Applied Physics Letters},
  title        = {Quantum-confined Stark effects in coupled InAs/GaAs quantum dots},
  doi          = {10.1063/1.4807770},
  issn         = {0003-6951},
  number       = {21},
  pages        = {213101},
  url          = {https://doi.org/10.1063/1.4807770},
  volume       = {102},
}

@Article{Bayer1998,
  author       = {Bayer, M. and Walck, S. N. and Reinecke, T. L. and Forchel, A.},
  title        = {Exciton binding energies and diamagnetic shifts in semiconductor quantum wires and quantum dots},
  pages        = {6584--6591},
  volume       = {57},
  date         = {1998-03},
  doi          = {10.1103/PhysRevB.57.6584},
  issue        = {11},
  journaltitle = {Phys. Rev. B},
  numpages     = {0},
  publisher    = {American Physical Society},
  url          = {https://link.aps.org/doi/10.1103/PhysRevB.57.6584},
}

@Article{Kuhlmann2013,
  author       = {Kuhlmann, Andreas V. and Houel, Julien and Ludwig, Arne and Greuter, Lukas and Reuter, Dirk and Wieck, Andreas D. and Poggio, Martino and Warburton, Richard J.},
  title        = {Charge noise and spin noise in a semiconductor quantum device},
  issn         = {1745-2481},
  number       = {9},
  pages        = {570--575},
  volume       = {9},
  date         = {2013},
  doi          = {10.1038/nphys2688},
  journaltitle = {Nature Physics},
  refid        = {Kuhlmann2013},
  url          = {https://doi.org/10.1038/nphys2688},
}

@Article{Abraham1999,
  author  = {Abraham-Shrauner, Barbara and Nordheden, Karen J. and Lee, Yao-Sheng},
  title   = {Model for etch depth dependence on GaAs via hole diameter},
  doi     = {10.1116/1.590677},
  issn    = {1071-1023},
  number  = {3},
  pages   = {961-964},
  url     = {https://doi.org/10.1116/1.590677},
  volume  = {17},
  journal = {Journal of Vacuum Science \& Technology B: Microelectronics and Nanometer Structures Processing, Measurement, and Phenomena},
  month   = {05},
  year    = {1999},
}

@Article{Lai2006,
  author   = {Lai, S. L. and Johnson, D. and Westerman, R.},
  title    = {Aspect ratio dependent etching lag reduction in deep silicon etch processes},
  doi      = {10.1116/1.2172944},
  issn     = {0734-2101},
  number   = {4},
  pages    = {1283-1288},
  url      = {https://doi.org/10.1116/1.2172944},
  volume   = {24},
  journal  = {Journal of Vacuum Science \& Technology A},
  month    = {06},
  year     = {2006},
}

@Article{McNevin1998,
  author   = {McNevin, S. C. and Cerullo, M.},
  title    = {Contact etch scaling with contact dimension},
  doi      = {10.1116/1.581179},
  issn     = {0734-2101},
  number   = {3},
  pages    = {1514-1518},
  url      = {https://doi.org/10.1116/1.581179},
  volume   = {16},
  journal  = {Journal of Vacuum Science \& Technology A},
  month    = {05},
  year     = {1998},
}

\end{document}